\documentclass[12pt]{article}
\usepackage{amsfonts}
\usepackage{amsmath}
\usepackage{amssymb,amsmath}
\usepackage{amscd}
\def\D{\Delta}

\def\cl{\centerline}

\def\vs{\vspace*}
\def\W{\mathcal {W}}

\def\U{\mathcal {U}}
\def\Z{\mathbb{Z}}

\def\F{\mathbb{F}}
\def\T{\mathcal {T}}
\def\QED{\hfill$\Box$}

\def\t{\tilde}
\def\wt{\widetilde}

\numberwithin{equation}{section}
\newtheorem{theo}{Theorem}[section]

\newtheorem{coro}[theo]{Corollary}
\newtheorem{lemm}[theo]{Lemma}
\newtheorem{prop}[theo]{Proposition}

\begin{document}

\cl{{\bf $q$-Deformation of $W(2,2)$ Lie algebra associated with
quantum groups}\footnote{Supported by
NSF grants 10825101 of China \\[2pt]\indent $^{*}$Corresponding author:
lmyuan@mail.ustc.edu.cn}}\vs{6pt}

\cl{Lamei Yuan$^{\dag}$}

\cl{\small $^{\dag}$Department of Mathematics, \!University of
Science and Technology \!of \!China, Hefei 230026, China}

\cl{\small E-mail: lmyuan@mail.ustc.edu.cn }\vs{6pt}

{\small\parskip .005 truein \baselineskip 3pt \lineskip 3pt

\noindent{{\bf Abstract.} An explicit
realization of the $W(2,2)$ Lie algebra is presented using the famous
 bosonic and fermionic  oscillators in physics, which is then used to construct the $q$-deformation
 of this Lie algebra. Furthermore, the quantum group structures on the $q$-deformation
 of this Lie algebra are
 completely determined.

 \vs{5pt}


\parskip .001 truein\baselineskip 6pt \lineskip 6pt

\vs{18pt}

\cl{\bf\S1. \
Introduction}\setcounter{section}{1}\setcounter{equation}{0}
\vs{8pt}

Since 1990s, there have been intensive explorations of quantized
universal enveloping algebras, namely, quantum groups,  which were
first introduced independently by Drinfeld \cite{D1,D2} and Jimbo
\cite{J1,J2} around 1985 in order for them to construct solutions to
the quantum Yang-Baxter equations. Since then quantum groups are
found to have numerous applications in various areas ranging from
statistical physics via symplectic geometry and knot theory to
modular representations of reductive algebraic groups. For this
reason, the interests in quantum groups, quantum deformations of Lie
algebras as well as Lie bialgebras  have been growing in the
physical and mathematical literatures, especially those of Cartan
type and Block type, which are closely related to the Virasoro
algebra and the $W$-infinity algebra ${\cal W}_{1+\infty}$ (e.g.,
\cite{LSX, P, SS, SSX, SSW, WSS, WSS2, XSS, YaS, YS}). In
particular, the $q$-deformed Virasoro algebra, $q$-deformed
oscillator and $q$-deformed Heisenberg algebra have been
investigated in a number of papers (see e.g. \cite{AS, CCSP, CILP,
H2, RM, OS}). Among these kinds of algebras, the $q$-deformation of
the Virasoro algebra has been most intensively considered \cite{AS,
FR, H, K, L, MT}, which can be viewed as a typical example of the
physical application of the quantum group. In addition,
two-parameter deformation of Lie algebras has also been considered
by some authors (see e.g. \cite{B,CJ}), while the more general
quantum Lie algebras have been investigated in \cite{CZ, DG, HLS,
SW}. Roughly  speaking, quantum Lie algebras in the context of these
deformations are  universal enveloping algebras deformed by one or
more parameter(s) ($q$-deformation) and possess structures of Hopf
algebras. However, the essential reason for the name ``quantum''
algebra is that it becomes the conventional Lie algebra in the
$q\rightarrow1$ limit (classical limit).

The $W(2,2)$ Lie algebra  was introduced by Zhang-Dong in \cite{ZD}
for the study of classification of vertex operator algebras
generated by vectors of weight $2$. Later the Harish-Chandra modules
of this Lie algebra were investigated in \cite{LZ}, while the
classification of irreducible weight modules was discussed in
\cite{LGZ}. The derivations, central extensions and automorphism
groups of this Lie algebra were determined in \cite{GJP}. Recently,
the Verma modules over the $W(2,2)$ Lie algebra was investigated in
\cite{JP}. A quantum group structure of the $q$-deformed
 $W(2,2)$ Lie algebra was also  given in \cite{FLC}.
 Nonetheless, there are still plenty rooms for our reconsideration on this matter,
since our definition of the $q$-deformation $\W_q$ (Proposition
\ref{prop1}) of this Lie algebra with its origin from physics, is rather different and thus the quantum group (associated with the deformation) constructed in this paper seems to be new.
\par

The $W(2,2)$ Lie algebra (denoted by $\mathcal {W}$) considered in
the present paper is an infinite dimensional Lie algebra with basis
$L_n$, $W_n$ (for $n\in\Z$) and the Lie brackets
\begin{eqnarray}\label{LE}
[L_m,L_n]=(n-m)L_{m+n},\ \ \ [L_m,W_n]=(n-m)W_{m+n},\ \ \
[W_m,W_n]=0, \ \ \ \forall\ \ m,n\in\Z.
\end{eqnarray}
One sees that it is different from that defined in \cite{ZD}, since
we drop here the central element. But our next aim is to develop the
central extensions of the $q$-deformation $\W_q$. \par

  In this paper, an explicit realization (Lemma \ref{lemm-real}) of the $W(2,2)$ Lie algebra defined in (\ref{LE})
is 
given using the famous  bosonic and fermionic oscillators
in physics ((\ref{boson}) and (\ref{fermion})).
As a result, the $q$-deformation $\W_q$ (Proposition \ref{prop1}) of the $W(2,2)$ Lie algebra is obtained by exact calculations. Based
on this,  Hopf algebraic structures of $\W_q$ are
proposed and thus the quantum group of $\W_q$ are completely
determined. It seems to us that the results in our paper may be of
some potential use in mathematical physics.

Throughout this paper, $\mathbb{F}$ denotes a field of
characteristic zero,  $\mathbb{F}^*$  the multiplicative group of
nonzero element of $\mathbb{F}$.  All vector spaces and algebras are
assumed to be over  $\mathbb{F}$. Denote by $\Z$, $\Z_+$, $\Z^*$ the
sets of integers, nonnegative, nonzero integers, respectively.
\vs{18pt}

\cl{\bf\S2. \ Realization of $\W$ and its $q$-Deformation $\W_q$
}\setcounter{section}{2}\setcounter{equation}{0} \vs{8pt}

 In this section, we propose a version of realizing the $\W(2,2)$
 Lie algebra defined in (\ref{LE}). Based on this, a quantum deformation of
 this Lie algebra is constructed. One sees that the $q$-deformed $\W(2,2)$ Lie algebra $\W_q$ here is rather different from that given in
 \cite{FLC}.\par

 In the oscillator, the bosonic oscillator $a$ and its
hermitian conjugate $a^+$ obey the commutation relations:
\begin{eqnarray}\label{boson}
[a,a^+]=aa^+-a^+a=1,\ \ \ [1,a^+]=[1,a]=0.
\end{eqnarray}
It follows by induction on $n$ that $[a,(a^+)^n]=n(a^+)^{n-1}$ for
all $n\in\Z.$ Then the generators
\begin{eqnarray}
L_n\equiv (a^+)^{n+1}a
\end{eqnarray}
realize the centerless Virasoro Lie algebra with bracket:
$$[L_m,L_n]=(n-m)L_{m+n}, \ \ \ \forall \ \ m,n\in\Z.$$ More details can be
consulted in \cite{L}.  \par For the realization of $W(2,2)$ Lie
algebra $\W$ defined in (\ref{LE}), in addition to the bosonic
oscillators $a$ and $a^{+}$, we introduce the fermionic oscillators
$b$ and $b^+$ with the anticommutators
\begin{eqnarray}\label{fermion}
\{b,b^+\}=bb^++b^+b=1;\ \ \ b^2=(b^+)^2=0.
\end{eqnarray}
Moreover, we set $[a,b]=[a,b^+]=[a^+,b]=[a^+,b^+]=0.$
 \begin{lemm}\label{lemm-real} With notations above, generators of the form
\begin{eqnarray}\label{real}
\ \ \ \ \ \ \ \ L_n\equiv (a^+)^{n+1}a;\ \ \ W_n\equiv
(a^+)^{n+1}b^+a,\ \ \ \forall\ n\in\Z,
\end{eqnarray}
realize the $W(2,2)$ Lie algebra $\W$ under the commutator
$$[A,B]=AB-BA,\ \ \ \forall\ \ A,B\in\W.$$
\end{lemm}
\noindent{\it Proof}.\ \ We have to check that $L_n$ and $W_n$
defined in (\ref{real}) satisfy the three relations in (\ref{LE}).
In fact, from $[a,(a^+)^n]=n(a^+)^{n-1}$ it follows
\begin{eqnarray*}
[L_n,L_m]&=&(a^+)^{n+1}a(a^+)^{m+1}a-(a^+)^{m+1}a(a^+)^{n+1}a\\
&=&(a^+)^{n+1}\Big((a^+)^{m+1}a+(m+1)(a^+)^m\Big)a\\
&\ &-(a^+)^{m+1}\Big((a^+)^{n+1}a+(n+1)(a^+)^{n}\Big)a\\
&=&(m-n)(a^+)^{m+n+1}a=(m-n)L_{m+n}.
\end{eqnarray*}
Similarly, one can get the other two equations,  namely,
$[L_n,W_m]=(m-n)W_{m+n}$ and $[W_n,W_m]=0$, since $b^+$ commutes
with both $a$ and $a^+$ and since $(b^+)^2=0.$\QED\vskip5pt\par Fix a
$q\in\F^*$ such that $q$ is not a root of unity. Instead of equation
(\ref{boson}), we assume that
\begin{eqnarray}\label{qboson}
[a,a^+]_{(q^{-1},q)}=1.
\end{eqnarray}
Here we use the notation:
\begin{eqnarray}\label{notation}
[A,B]_{(\alpha,\beta)}=\alpha AB-\beta BA.
\end{eqnarray}
From(\ref{qboson}) it follows by induction on $n$ that
\begin{eqnarray}\label{qd}
[a,(a^+)^n]_{(q^{-n},q^n)}=[n]_q(a^+)^{n-1} \end{eqnarray}
for
arbitrary $n$, where the general notation
$$[n]_q=\frac{q^n-q^{-n}}{q-q^{-1}}$$
is used. It is clear to see that $[-n]_q=-[n]_q.$ Furthermore, one
can also deduce that
\begin{eqnarray}\label{q}
q^{n}[m]_q-q^{m}[n]_q=[m-n]_q, \ \ \ q^{-n}[m]_q+q^{m}[n]_q=[m+n]_q.
\end{eqnarray}
\par Now we have the
following result:
\begin{prop}\label{prop1} The generators $L_n$ and $W_n$ (for
$n\in\Z$) defined in (\ref{real}) satisfy the following relations:
\begin{eqnarray}
&&[L_n,L_m]_{(q^{n-m},\,q^{m-n})}=[m-n]_qL_{m+n},\label{qLE1}\\
&&[L_n,W_m]_{(q^{n-m},\,q^{m-n})}=[m-n]_qW_{m+n},\label{qLE2}\\
&&[W_n,W_m]_{(q^{n-m},q^{m-n})}=0\label{qLE3},
\end{eqnarray}
for all $m,n\in\Z.$
\end{prop}
\noindent{\it Proof}.\ \ Obviously, equation (\ref{qLE1}) holds for
$m=n$ since both sides are equal to $0$. Assume that $n\neq m$. Then
it follows from (\ref{real}), (\ref{notation}), (\ref{qd}) and
(\ref{q}) that
\begin{eqnarray*}
[L_n,L_m]_{(q^{n-m},q^{m-n})}&=&q^{n-m}(a^+)^{n+1}a(a^+)^{m+1}a-q^{m-n}(a^+)^{m+1}a(a^+)^{n+1}a\\
&=&q^{n+1}(a^+)^{n+1}\Big(q^{-m-1}a(a^+)^{m+1}\Big)a-q^{m+1}(a^+)^{m+1}\Big(q^{-n-1}a(a^+)^{n+1}\Big)a\\
&=&q^{n+1}(a^+)^{n+1}\Big(q^{m+1}(a^+)^{m+1}a+[m+1]_q(a^+)^m\Big)a\\
&\ &-q^{m+1}(a^+)^{m+1}\Big(q^{n+1}(a^+)^{n+1}a+[n+1]_q(a^+)^n\Big)a\\
&=&\Big(q^{n+1}[m+1]_q-q^{m+1}[n+1]_q\Big)(a^+)^{m+n+1}a\\
&=&[m-n]_q(a^+)^{m+n+1}a=[m-n]_qL_{m+n}.
\end{eqnarray*}
Hence equation (\ref{qLE1}) holds for all $m,n\in\Z.$ Similarly, one
can get  equations (\ref{qLE2}) and (\ref{qLE3}) using the facts
that $b^+$ commutes with both $a$ and $a^+$ and that
$(b^+)^2=0.$\QED\vskip5pt\par

Proposition \ref{prop1}  says that the algebra with generators
$L_n$, $W_n$ (for $n\in\Z$) and the relations
(\ref{qLE1})--(\ref{qLE3}) realizes  the $W(2,2)$ Lie algebra
$\W$ in the $q\rightarrow1$ limit. We call it the {\it $q$-deformation} of
the $W(2,2)$ Lie algebra $\W$, which will be denoted by $\W_q$ in
the sequel.

Furthermore, we can generalize this deformation by setting
\begin{eqnarray*}
[a,a^+]_{(q^c,q)}=1,
\end{eqnarray*}
where $c\in\F^*$ and $c\neq 1$. Then by induction, one has
\begin{eqnarray*}
[a,(a^+)^n]_{(q^{nc},q^n)}=[n]_q^c(a^+)^{n-1},
\end{eqnarray*}
where
\begin{eqnarray*}
[n]_{q}^c=\frac{q^n-q^{nc}}{q-q^c}.
\end{eqnarray*}
Now we can show that the expressions  (\ref{real}) of $L_n$ and
$W_n$ satisfy the following relations:
\begin{eqnarray}
&&[L_n,L_m]_{(q^{n-m},\,q^{c(n-m)})}=-[n-m]_q^cL_{m+n},\label{gq1}\\
&&[L_n,W_m]_{(q^{n-m},\,q^{c(n-m)})}=-[n-m]_q^cW_{m+n},\label{gq2}\\
&&[W_n,W_m]_{(q^{n-m},\,q^{c(n-m)})}=0\label{gq3},
\end{eqnarray}
for all $m,n\in\Z$. From these we see that the algebra generated by
$L_n$ and $W_n$ in (\ref{real}) with relations
(\ref{gq1})--(\ref{gq3}) is also a realization of the $W(2,2)$ Lie
algebra when $q\rightarrow 1$, which is called the generalized
$q$-deformation of $W(2,2)$ Lie algebra and is denoted by $\W_q^c$.
Note that one gets $\W_q$ (see Proposition \ref{prop1}) when $c=-1.$

\vs{18pt}

\cl{\bf\S3. \ Quantum Group Structures of $\W_q$
}\setcounter{section}{3}\setcounter{equation}{0} \vs{8pt}

In this section, we give a direct construction of the Hopf algebraic
structures of the $q$-deformed $W(2,2)$ Lie algebra $\W_q$ defined
in previous section.\par Fix a $q\in\F^*$ such that $q$ is not a
root of unity. Then $\mathcal {U}_q$ is defined as the associative
algebra (with $1$ and over $\F$) with generators $\mathcal {T}$,
$\mathcal {T}^{-1}$, $L_n$,$W_n$ for $n\in\Z$ and relations:

\begin{eqnarray*}
(R1)\ \ \ \ \ \ \ \ \ \ \ \ \ \ \ \ \ \ &&\T\T^{-1}=1=\T^{-1}\T;\ \ \ \ \ \ \ \ \ \ \ \ \ \ \ \ \ \ \ \ \\
(R2)\ \ \ \ \ \ \ \ \ \ \ \ \ \ \ \ \ \ &&\T^mL_n=q^{-2(n+1)m}L_n\T^m;\ \ \ \ \ \ \ \ \ \ \ \ \ \ \ \  \\
(R3)\ \ \ \ \ \ \ \ \ \ \ \ \ \ \ \ \ \
&&\T^mW_n=q^{-2(n+1)m}W_n\T^m;\ \ \ \ \ \ \ \ \ \ \ \
\ \ \ \ \ \ \ \ \ \ \ \\
(R4)\ \ \ \ \ \ \ \ \ \ \ \ \ \ \ \ \ \ &&
q^{n-m}L_nL_m-q^{m-n}L_mL_n=[m-n]_qL_{m+n};\ \ \
\ \ \ \ \ \ \ \ \ \ \ \ \ \ \ \ \\
(R5)\ \ \ \ \ \ \ \ \ \ \ \ \ \ \ \ \ \ &&
q^{n-m}L_nW_m-q^{m-n}W_mL_n=[m-n]_qW_{m+n};\ \ \ \ \ \ \ \ \
\ \ \ \ \ \ \ \ \ \ \ \ \ \ \ \ \ \ \ \ \ \ \\
(R6)\ \ \ \ \ \ \ \ \ \ \ \ \ \ \ \ \ \ &&
q^{n-m}W_nW_m-q^{m-n}W_mW_n=0.
\end{eqnarray*}
\par

Before giving the construction of the Hopf algebraic structures on
$\U_q$, we have to check whether  or not these six relations $(R1)$--$(R6)$
above ensure a nontrivial associative algebra $\U_q$. The
following proposition gives a positive answer. \par
\begin{prop} The associative algebra $\U_q$ with generators $\T$,
$\T^{-1}$, $L_n$, $W_n$ ($n\in\Z$) and relations $(R1)$--$(R6)$ is
nontrivial.
\end{prop}
\noindent{\it Proof}.\ \ Set $M:=\{L_n,\, M_n,\, \T,\,
\T^{-1}\big|\, n\in\Z\}$. Let $T(M)$ be the tensor algebra of $M$,
which is a free associative algebra generated by $M.$ Then one has
$$T(M)=\mbox{$\bigoplus\limits_{m=0}^{\infty}$} T(M)_m\,,$$
where $T(M)_m=\overbrace{M\otimes\cdots\otimes
M}^m=\mbox{span}\{v_1\otimes v_2\otimes \cdots\otimes
v_m\big|\,v_i\in M, \,i=1,2,\cdots,m\} $. In particular, $T(M)_0=\F$
and $T(M)_1=M.$ The product on $T(M)$ is naturally defined by
$$(v_1\otimes
v_2\otimes \cdots\otimes v_m)(w_1\otimes w_2\otimes\cdots\otimes
w_n)=v_1\otimes v_2\otimes \cdots\otimes v_m\otimes w_1\otimes
w_2\otimes\cdots\otimes w_n. $$ Let $I$ be the two--sided ideal of
$T(M)$ generated by
\begin{eqnarray}
&&\T\otimes\T^{-1}-\T^{-1}\otimes\T,\ \ \ \ \ \  \ \ \ \ \ \ \ \ \ \
q^{n-m}W_n\otimes
W_m-q^{m-n}W_m\otimes W_n;\label{I1}\\
&&\T^m\otimes L_n-q^{-2(n+1)m}L_n\otimes\T^m,\ \ \ \T^m\otimes
W_n-q^{-2(n+1)m}W_n\otimes\T^m
;\label{I2}\\
&&q^{n-m}L_n\otimes L_m-q^{m-n}L_m\otimes L_n,\ \ \ q^{n-m}L_n\otimes
W_m-q^{m-n}W_m\otimes L_n,\label{I3}
\end{eqnarray}
for all $m,n\in\Z$ and where $\T^{-n}=(\T^{-1})^{n}.$ Set
$S(M):=T(M)/I$. It is obvious that $S(M)$ is also a $\Z$--graded
associative algebra with a basis
\begin{eqnarray}\label{basis}
\t{B}=\{T^d (\T^{-1})^{d'}L_{i_1}^{k_1}\cdots
L_{i_m}^{k_m}W_{j_1}^{l_1}\cdots W_{j_n}^{l_n}\},
\end{eqnarray}
where $d,\,d',\, k_i,\, l_j,\, i_p,\, j_q\in\Z_+,\ ( i,p=1,2,\cdots
m; \ j,q=1,2,\cdots n); \ i_1<i_2<\cdots <i_m,\ j_1<j_2<\cdots
<j_n$. Let $\tilde{J}$ be another two--sided ideal of $T(M)$
generated by the elements of form
\begin{eqnarray}
&&q^{n-m}L_n\otimes
L_m-q^{m-n}L_m\otimes L_n-[L_n,L_m]_{(q^{n-m},\,q^{m-n})},\label{J3}\\
&&q^{n-m}L_n\otimes W_m-q^{m-n}W_m\otimes
L_n-[L_n,W_m]_{(q^{n-m},q^{m-n})},\label{J4}
\end{eqnarray}
together with that in (\ref{I1}) and (\ref{I2}). Then set
$$\widetilde{\U_q}:=T(M)/\t{J}.$$

Our  aim is to show that $\t{B}$ defined in (\ref{basis}) is
also a basis of $\widetilde{\U_q}$. Let
$$\t{B'}=\{v_{i_1}\otimes v_{i_2}\otimes\cdots\otimes v_{i_m}|v_i\in M, 1\leq i_1\leq i_2 \leq \cdots \leq i_m, m\geq0\}$$
be a subset of $T(M)$ and let $U'$ be the subspace of $T(M)$ spanned
by $\t{B'}$. We now claim that
\begin{eqnarray}\label{sum}
T(M)=U'\mbox{$\bigoplus$}\,\t{J}.
\end{eqnarray}For any $v\in
T(M)$, we can write $v$ as
 $v=v^{(m)}+v^{(m-1)}+\cdots +v^{0}$,  where $v^{(m)}\neq 0$ for some $m\geq 0$ and where
$v^{(i)}\in T(M)_i$ with $i=0,1,\cdots,m.$ We call $m$ the {\it
degree} of $v$. From (\ref{I1}), (\ref{I2}),
(\ref{J3}) and (\ref{J4}), it follows
\begin{eqnarray*}
v_{i_1}\otimes \cdots\otimes \Big(v_{i_k}\otimes
v_{i_{k+1}}-v_{i_{k+1}}\otimes
v_{i_{k}}-[v_{i_k},\,v_{i_{k+1}}]\Big)\otimes \cdots \otimes
v_{i_m}\in\t{J},
\end{eqnarray*}
namely, the difference between $v_{i_1}\otimes \cdots\otimes
v_{i_k}\otimes v_{i_{k+1}}\otimes \cdots \otimes v_{i_m}$ and $a
v_{i_1}\otimes \cdots\otimes v_{i_{k+1}}\otimes v_{i_{k}}\otimes
\cdots \otimes v_{i_m}$ (for some $a\in\F^*$) is an element in
$\t{J}$ and an element with degree less than $m$. So by induction on
the degree of $v$ one can obtain that $T(M)=U'+\t{J}$.\par

 It remains to show that equation (\ref{sum}) is a direct sum, which is equivalent to the linear independence of $\t{B}$ in $\wt{U_q}$.
 Suppose that a nonzero linear combination $v$ of the elements in
 $\t{B'}$ is in $\t{J}$. It follows from (\ref{I1}), (\ref{I2}), (\ref{J3}) and (\ref{J4})
that the homogeneous component $v^{(m)}$ of $v$ with highest degree
must lie in $\ker\pi$ (by comparing (\ref{I3}) with (\ref{J3}) and
(\ref{J4})), where $\pi:T(M)\rightarrow S(M)$ is the natural
$\Z$--graded algebraic homomorphism, namely,
$$\pi(v_{i_1}\otimes v_{i_2}\otimes \cdots \otimes v_{i_m})=v_{i_1}v_{i_2}\cdots v_{i_m}.$$
However, $v^{(m)}$ is a nonzero linear combination of the elements
in $\t{B'}$, it is impossible to appear in $\ker\pi$. This
contradiction implies $\t{B}$ is a basis of $\wt{\U_q}$. Since it is
clear that $\U_q\cong\wt{\U_q}/J$, where $J$ is the two-sided ideal
of $\wt{\U_q}$ generated by $\T\T^{-1}-1$, we obtain that $\U_q$ is
a nontrivial associative algebra with basis
\begin{eqnarray}
\t{B}=\{T^d L_{i_1}^{k_1}\cdots L_{i_m}^{k_m}W_{j_1}^{l_1}\cdots
W_{j_n}^{l_n}\},
\end{eqnarray}
where $d\in\Z,\, k_i,\, l_j,\, i_p,\, j_q\in\Z_+,\ ( i,p=1,2,\cdots
m; \ j,q=1,2,\cdots n); \ i_1<i_2<\cdots <i_m,\ j_1<j_2<\cdots
<j_n$.\QED\vskip5pt\par With the above proposition in hand, we can safely
proceed with the construction of the Hopf algebraic structures on
$\U_q$ now. This will be done by several lemmas below.
\begin{lemm}\label{lemm1} There is a unique homomorphism of
$\,\F$-algebras $\Delta:\U_q\rightarrow\U_q\times\U_q$ with
\begin{eqnarray}
&&\D(\T)=\T\otimes\T,\label{det1}\ \ \ \ \ \ \ \ \ \ \ \ \ \ \ \ \ \ \ \ \ \ \ \ \\
&&\D(\T^{-1})=\T^{-1}\otimes\T^{-1},\label{det2}\\
&&\D(L_n)=L_n\otimes\T^n+\T^n\otimes L_n,\label{det3}\ \ \ \ \ \ \ \ \ \ \ \ \\
&&\D(W_n)=W_n\otimes\T^n+\T^n\otimes W_n.\label{det4}
\end{eqnarray}
\end{lemm}
\noindent{\it Proof}.\ \  It is clear that $\D(\T^{m})=\T^{m}\otimes
\T^{m}$ for arbitrary $m\in\Z.$ We have to show that $\D(\T)$, $
\D(\T^{-1})$, $ \D(L_n)$ and  $\D(W_n)$ satisfy the relations
($R1$)--($R6$). This is trivial for ($R1$). For ($R2$) and ($R3$) it
follows directly from (\ref{det1})--(\ref{det4}). Now look at
($R4$): we have
\begin{eqnarray*}
\D(L_n)\D(\L_m)&=&(L_n\otimes\T^n+\T^n\otimes
L_n)(L_m\otimes\T^m+\T^m\otimes L_m)\\
&=&L_nL_m\otimes\T^{m+n}+L_n\T^m\otimes\T^nL_m+\T^nL_m\otimes
L_n\T^m+\T^{m+n}\otimes L_nL_m\\
&=&L_nL_m\otimes\T^{m+n}+q^{-2(m+1)n}L_n\T^m\otimes
L_m\T^n\\
&\ &+\T^{m+n}\otimes L_nL_m +q^{-2(m+1)n}L_m\T^n\otimes
L_n\T^m.\\
\end{eqnarray*}
Similarly, we get

\begin{eqnarray*}
\D(L_m)\D(\L_n)&=&L_mL_n\otimes\T^{m+n}+q^{-2(n+1)m}L_m\T^n\otimes
L_n\T^m\\
&\ &+\T^{m+n}\otimes L_mL_n +q^{-2(n+1)m}L_n\T^m\otimes
L_m\T^n.\\
\end{eqnarray*}
Then it follows
\begin{eqnarray*}
&&\ \ q^{n-m}\D(L_n)\D(\L_m)-q^{m-n}\D(L_m)\D(\L_n)\\
&&=\big(q^{n-m}L_nL_m\otimes\T^{m+n}-q^{m-n}L_mL_n\otimes\T^{m+n}\big)
+\big(q^{n-m}\T^{m+n}\otimes L_nL_m-q^{m-n}\T^{m+n}\otimes
L_mL_n\big)\\
&&=[n-m]_q\big(L_{m+n}\otimes \T^{m+n}+\T^{m+n}\otimes
L_{m+n}\big)=[n-m]_q\D(L_{m+n}).
\end{eqnarray*}
Hence, ($R4$) is preserved by $\D$ and so is it  for ($R5$) and
($R6$), which can be checked by the similar method. That means $\D$
is an algebraic homomorphism. Consequently, $\U_q$ is a
bialgebra.\QED\vskip4pt\par The map $\D$ from Lemma \ref{lemm1} is called the
{\it comultiplication} on $\U_q$. We say $\D$ is {\it
coassociative}, if it satisfies $(1\otimes\D)\D=(\D\otimes1)\D$.
\begin{lemm}\label{lemm2}
The comultiplication $\D$ on $\U_q$ is coassociative.
\end{lemm}
\noindent{\it Proof}.\ \ We simply have to check that all the
generators of $\U_q$ are mapped both ways by $(1\otimes\D)\D$ and
$(\D\otimes1)\D$ to the same image, which simply involves straightforward calculations. We
shall take $L_n$  as an example (others can be done similarly).
\begin{eqnarray*}
(1\otimes\D)\D(L_n)&=&(1\otimes\D)(L_n\otimes\T^n+\T^n\otimes
L_n)=L_n\otimes\D(\T^n)+\T^n\otimes\D(L_n)\\
&=&L_n\otimes(\T^n\otimes\T^n)+\T^n\otimes(L_n\otimes\T^n+\T^n\otimes
L_n)\\
&=&(L_n\otimes\T^n)\otimes\T^n+(\T^n\otimes
L_n)\otimes\T^n+(\T^n\otimes\T^n)\otimes
L_n\\
&=&\D(L_n)\otimes\T^n+\D(\T^n)\otimes L_n=(\D\otimes
1)(L_n\otimes\T^n+\T^n\otimes L_n)\\&=&(\D\otimes1)\D(L_n).
\end{eqnarray*}
\vspace*{-24pt}

\QED\par
\begin{lemm}\label{lemm3}
There is a unique homomorphism of $\F$-algebras $\varepsilon:\U_q
\rightarrow \F $ with
\begin{eqnarray}\label{eps}
\varepsilon(\T)=\varepsilon(\T^{-1})=1 \ \ \mbox{and} \ \
\varepsilon(L_n)=\varepsilon(W_n)=0,
\end{eqnarray}
for all $n\in\Z$. Moreover, the following diagrams are commutative
\[\begin{CD}
\U_q@>\D>>\U_q\otimes\U_q\\
@V{\rm id} VV  @VV1\otimes \varepsilon V\\
\U_q@>>\pi_1>\U_q\otimes\U_q\\
\end{CD}\ \  \   \ \ \ \ \ \ \ \ \ \ \ \ \
\begin{CD}
\U_q@>\D>>\U_q\otimes\U_q\\
@V{\rm id}VV  @VV \varepsilon\otimes1 V\\
\U_q@>>\pi_2>\U_q\otimes\U_q\\
\end{CD}\]
namely, $(1\otimes\varepsilon)\Delta=\pi_1\circ{\rm id}$ and
$(\varepsilon\otimes 1)\Delta=\pi_2\circ{\rm id}$, where $\pi_1$
(resp. $\pi_2$) denotes the isomorphism $u\mapsto u\otimes 1$ (resp.
$u\mapsto 1\otimes u$) for any $u\in\U_q.$
\end{lemm}
\noindent{\it Proof}.\ \ It is straightforward to see that
$\big(\varepsilon(\T), \varepsilon(\T^{-1}), \varepsilon(L_n),
\varepsilon(W_n)\big)$=($1,1,0,0$) satisfy the relations
\mbox{$(R1)$--$(R6)$}. So we have the homomorphism $\varepsilon$.
For the commutativity of the diagrams, it can be easily checked on
the generators.\QED\vskip5pt\par The homomorphism $\varepsilon$ from
Lemma \ref{lemm3} is called the {\it counit} of $\U_q.$
\begin{lemm}\label{lemm4}
There is a unique antiautomorphism $S$ of $\U_q$ with
\begin{eqnarray}
&&S(L_m)=-\T^{-m}L_m\T^{-m},\label{antipode1}\\
&&S(W_m)=-\T^{-m}W_m\T^{-m},\label{antipode2}\\
&&S(\T)=\T^{-1}, \ S(\T^{-1})=\T.\label{antipode3}
\end{eqnarray}
In addition, one has $S^2={\rm id}.$
\end{lemm}
\noindent{\it Proof}.\ \ We need to show that
$\big(S(\T),S(\T^{-1}), S(L_m), S(W_m)\big)$ satisfy the relations
$(R1)$--$(R6)$ in $\U_q^{opp}$. Let us denote the multiplication in
$\U_q^{opp}$ by a ``\,$\cdot$\,'' in order  to distinguish it from that in $\U_q.$
Now $(R1)$ is obvious and it is easy to see that $S(\T^{m})=\T^{-m}$
$(m\in\Z)$. For $(R2)$ we have
\begin{eqnarray*}
S(\T^m)\cdot
S(L_m)&=&S(L_m)S(\T^m)=-\T^{-m}L_m\T^{-m}\T^{-m}=-\T^{-m}q^{-2(m+1)m}\T^{-m}L_m\T^{-m}\\
&=&q^{-2(m+1)m}S(\T^m)S(L_m)=q^{-2(m+1)m}S(L_m)\cdot S(\T^m).
\end{eqnarray*}
One can check similarly that $(R3)$ is preserved by $S$. Now consider $(R4)$: it follows from
$(R2)$ that
\begin{eqnarray*}
q^{n-m}S(L_n)\cdot
S(L_m)&=&q^{n-m}S(L_m)S(L_n)=q^{n-m}\T^{-m}L_m\T^{-m}\T^{-n}L_n\T^{-n}\\
&=&q^{n-m}q^{-2(m+1)n}\T^{-m-n}L_mq^{2(n+1)m}L_n\T^{-m-n}\\
&=&q^{m-n}\T^{-m-n}L_mL_n\T^{-m-n}.
\end{eqnarray*}
Similarly, one has $q^{m-n}S(L_m)\cdot
S(L_n)=q^{n-m}\T^{-m-n}L_nL_m\T^{-m-n}$. Then we have
\begin{eqnarray*}
q^{n-m}S(L_n)\cdot S(L_m)-q^{m-n}S(L_m)\cdot
S(L_n)&=&\T^{-m-n}\big(q^{m-n}L_mL_n-q^{n-m}L_nL_m\big)\T^{-m-n}\\
&=&-[m-n]_q \T^{-m-n}L_{m+n}\T^{-m-n}\\
&=&[m-n]_qS(L_{m+n}),
\end{eqnarray*}
namely, the map $S$ preserves $(R4)$. One can similarly check that
$(R5)$ and  $(R6)$ are also preserved by $S.$ So there is indeed a
homomorphism $S:\U_q\rightarrow \U_q^{opp}$ or an antihomomorphism
$S:\U_q\rightarrow \U_q$ satisfying
(\ref{antipode1})-(\ref{antipode3}). Now $S^2$ is an ordinary
homomorphism from $\U_q$ to $\U_q$. One can check easily on the
generators that $S^2=\rm id$, which implies that $S$ is
bijective.\QED\vskip5pt\par The map $S$ in Lemma \ref{lemm4} is
called the {\it antipode} of $\U_q$. It is clear that the inverse
$S^{-1}$
 of $S$ is an antiautomorphism, which is given by
$$S^{-1}(\T)=\T^{-1},\ \ S^{-1}(L_m)=-\T^{-m}L_m\T^{-m}, \ \ \  S^{-1}(W_m)=-\T^{-m}W_m\T^{-m},\ \ \forall\ \ m\in\Z.$$
\begin{lemm}\label{lemm5}
The following diagrams are commutative
\[\begin{CD}
\U_q@>\D>>\U_q\otimes\U_q\\
@V\iota\circ\varepsilon VV  @VV1\otimes SV\\
\U_q@<m<<\U_q\otimes\U_q\\
\end{CD}\ \  \   \ \ \ \ \ \ \ \ \ \ \ \ \
\begin{CD}
\U_q@>\D>>\U_q\otimes\U_q\\
@V\iota\circ\varepsilon VV  @VV S\otimes 1V\\
\U_q@<m<<\U_q\otimes\U_q\\
\end{CD}\]
where $m:\U_q\otimes\U_q\rightarrow\U_q$ is the multiplication map,
namely, $m(u\otimes u')=uu'$ for all $u,u'\in\U_q$, and where
$\iota:\F\rightarrow\U_q$ is the embedding $\iota(a)=a1$ for all
$a\in\F$.
\end{lemm}
\noindent{\it Proof}.\ \ Let us restrict ourselves to the left
diagram. The map $f=m\circ(1\otimes S)\circ\D$ acts on
generators as follows:
\begin{eqnarray*}\!\!\!\!\!\!\!\!\!\!\!\!
&&\T\mapsto\T\otimes\T\mapsto\T\otimes\T^{-1}\mapsto\T\T^{-1}=1,\\
\!\!\!\!\!\!\!\!\!\!\!\!&&\T^{-1}\mapsto\T^{-1}\otimes\T^{-1}\mapsto\T^{-1}\otimes\T\mapsto\T^{-1}\T=1,\\
\!\!\!\!\!\!\!\!\!\!\!\!&&L_n\mapsto L_n\otimes\T^n+\T^{n}\otimes L_n\mapsto L_n\otimes
\T^{-n}+\T^{n}\otimes(-\T^{-n}L_n\T^{-n})\mapsto
L_n\T^{-n}-L_n\T^{-n}=0,\\
\!\!\!\!\!\!\!\!\!\!\!\!&&W_n\mapsto W_n\otimes\T^n+\T^{n}\otimes W_n\mapsto W_n\otimes
\T^{-n}+\T^{n}\otimes(-\T^{-n}W_n\T^{-n})\mapsto
W_n\T^{-n}-W_n\T^{-n}=0,
\end{eqnarray*}
as predicted by the diagram.\par To conclude the proof we have to
check: If $f(u)=\iota\circ \varepsilon(u)$ and $f(v)=\iota\circ
\varepsilon(v)$ for $u,v\in \U_q$, then also $f(uv)=\iota\circ
\varepsilon(uv).$ That is not obvious, since $S$ and $m$ are not
ring homomorphisms. We suppose that $\D(u)=\mbox{$\sum_i$}u_i\otimes
u_i'$ and $\D(v)=\mbox{$\sum_i$}v_i\otimes v_i'$ in
$\U_q\otimes\U_q.$ Then $f(uv)$ is given by
\begin{eqnarray*}
uv\mapsto \mbox{$\sum\limits_{i,j}$}u_iv_j\otimes
u_i'v_j'\mapsto\mbox{$\sum\limits_{i,j}$}u_iv_j\otimes
S(v_j')S(u_i')\mapsto
\mbox{$\sum\limits_{i,j}$}u_iv_jS(v_j')S(u_i')\mapsto
\mbox{$\sum\limits_i$}u_if(v)S(u_i'),
\end{eqnarray*}
since
\begin{eqnarray*}
f(v)=m\circ(1\otimes S)\circ\D(v)=m\circ(1\otimes
S)(\mbox{$\sum_j$}v_j\otimes v_j')=m(\mbox{$\sum_j$}v_j\otimes
S(v_j'))=\mbox{$\sum_j$}v_jS(v_j').
\end{eqnarray*}
We assume that $f(v)=\iota\circ\varepsilon(v)$, so this element is a
scalar multiple of $1$ and thus central in $U_q$. Therefore
$$f(uv)=\mbox{$\sum\limits_i$}u_iS(u_i')f(v)=f(u)f(v)=\iota\circ\varepsilon(uv).\eqno
\mbox{\QED}$$
\par
In general, an $\F$-algebra $A$ together with algebra homomorphisms
$\D:A\rightarrow A\otimes A$ and $\varepsilon:A\rightarrow\F$ and a
linear map $S:A\rightarrow A$ is called a {\it Hopf algebra}, if
$\D$ is coassociative and if the diagrams in Lemmas \ref{lemm3} and
\ref{lemm5} (with $\U_q$ is replaced by $A$) commute. One calls
$\D$ the {\it comultiplication}, $\varepsilon$ the {\it counit} and
$S$ the {\it antipode} of the Hopf algebra. A Hopf algebra $A$ is
called {\it cocommutative}, if $P\circ\D=\D$ with $P(u\otimes
v)=v\otimes u$ for all $u$ and $v$ in $A$. So the Lemmas
\ref{lemm1}--\ref{lemm5} say:
\begin{theo}$(\U_q,\D,\varepsilon,S)$ defined by $(R1)$--$(R6)$ and \eqref{det1}--\eqref{antipode3} is a Hopf
algebra, which is neither cocommutative nor commutative.
\end{theo}\par
\begin{coro} As vector spaces, one has
$$\U_q\cong \F[\T,\T^{-1}]\otimes_{\F}U_q,$$
where $U_q=U(\W_q)$ is the enveloping algebra of $\W_q$ generated by
$L_n$ and $ W_n$ $(n\in\Z)$ with relations
\eqref{qLE1}--\eqref{qLE3}.
\end{coro}
\begin{coro}
\begin{eqnarray}
\!\!\!\!\!\!\!\!\!\!\!\!&&\D(\T^r)=\T^r\otimes\T^r,\ \ \ \ \ \ \ \ \ \ \ \ \ \ \ \ \ \ \ \ \ \ \ \ \ \ \ \ \ S(\T^r)=\T^{-r},\ \ \ \forall\ \ r\in\Z\,;\label{sd000}\\
\!\!\!\!\!\!\!\!\!\!\!\!&&\D(L_n^r)=\mbox{$\sum\limits_{i=0}^r$}{r\choose
i} L_n^{r-i}\T^{in}\otimes \T^{(r-i)n}L_n^i,\ \ \ \ \
S(L_n^r)=(-1)^r\T^{-rn}L_n^r\T^{-rn},\  \forall\ \ r\in\Z_+\,;\label{ds1}\\
\!\!\!\!\!\!\!\!\!\!\!\!&&\D(W_n^r)=\mbox{$\sum\limits_{i=0}^r$}{r\choose
i} W_n^{r-i}\T^{in}\otimes \T^{(r-i)n}W_n^i,\ \ \
S(W_n^r)=(-1)^r\T^{-rn}W_n^r\T^{-rn},\ \forall\ \
r\in\Z_+\,;\label{ds2}
\end{eqnarray}
for any $n\in\Z.$
\end{coro}
\noindent{\it Proof.~}~Equations in \eqref{sd000} are easily
obtained from (\ref{det1}) and (\ref{antipode3}).
 One sees that the formulas in
(\ref{ds1}) holds trivially for $r=0$, that is, $\D(1)=1\otimes 1$ and $S(1)=1$.
Using  definitions in Lemmas \ref{lemm1} and
\ref{lemm4}, one sees that (\ref{ds1}) holds for $r=1$. Here are then the inductive steps:
\begin{eqnarray*}
\D(L_n^{r+1})&=&(L_n\otimes\T^n+\T^n\otimes
L_n)\Big(\mbox{$\sum\limits_{i=0}^{r}$}{r\choose
i}L_n^{r-i}\T^{in}\otimes\T^{(r-i)n}L_n^i\Big)\\
&=&\mbox{$\sum\limits_{i=0}^r$}{r\choose
i}\Big(L_n^{r+1-i}\T^{in}\otimes
\T^{(r+1-i)n}L_n^i+\T^nL_n^{r-i}\T^{in}\otimes
L_n\T^{(r-i)n}L_n^i\Big)\\
&=&\mbox{$\sum\limits_{i=0}^r$}{r\choose
i}\Big(L_n^{r+1-i}\T^{in}\otimes
\T^{(r+1-i)n}L_n^i+L_n^{r-i}\T^{(i+1)n}\otimes
\T^{(r-i)n}L_n^{i+1}\Big)\\
&=&\mbox{$\sum\limits_{i=0}^r$}{r\choose
i}\Big(L_n^{r+1-i}\T^{in}\otimes
\T^{(r+1-i)n}L_n^i\Big)+\mbox{$\sum\limits_{i=1}^{r+1}$}{r\choose
{i-1}}\Big(L_n^{r+1-i}\T^{in}\otimes
\T^{(r+1-i)n}L_n^{i}\Big)\\
&=&\mbox{$\sum\limits_{i=0}^{r+1}$}\Big({r\choose
i}+{r\choose{i-1}}\Big)L_n^{r+1-i}\T^{in}\otimes
\T^{(r+1-i)n}L^i_n\\
&=&\mbox{$\sum\limits_{i=0}^{r+1}$}{{r+1}\choose
i}L_n^{r+1-i}\T^{in}\otimes \T^{(r+1-i)n}L^i_n,
\end{eqnarray*}
and
\begin{eqnarray*}
S(L_n^{r+1})&=&(-1)^r\T^{-rn}L_n^r\T^{-rn}\big(-\T^{-n}L_n\T^{-n}\big)=(-1)^{r+1}\T^{-rn}\big(L_n^r\T^{-n}\big)\big(\T^{-rn}L_n\big)\T^{-n}\\
&=&(-1)^{r+1}\T^{-rn}\big(q^{-2rn(n+1)}\T^{-n}L_n^r\big)\big(q^{2rn(n+1)}L_n\T^{-rn}\big)\T^{-n}\\
&=&(-1)^{r+1}\T^{-(r+1)n}L_n^{r+1}\T^{-(r+1)n}.
\end{eqnarray*}
Hence equations in (\ref{ds1}) hold by induction. Equations in
(\ref{ds2}) can be similarly proved.\QED


\vskip10pt\small\footnotesize

\begin{thebibliography}{9999}\vskip0pt
\parindent=2ex\parskip=-1pt\baselineskip=-1pt
\def\RE#1{\bibitem{#1}\label{#1}}

\RE{AS}N. Aizawa, H. Sato, $q$-Deformation of the Virasoro algebra
with central extension, {\it Phys. Lett. B}, {\bf 256}(2) (1991),
185--190.

\RE{B} I. Burban, Two-parameter deformation of the oscillator
algebra and $(p,q)$-analog of two-dimensional conformal field
theory, {\it Nonlinear Math. Phys.}, {\bf2}(1995), 384--391.

\RE{CCSP} K. H. Cho, R. Chaiho, D. S. Soh, S. U. Park, $q$-deformed
oscillator associated with the Calogero model and its $q$-coherent
state, {\it J. Phys. A: Math. Gen.}, {\bf 27}(1994) 2811--2822.

\RE{CILP}  M. Chaichian, A. P. Isaev, J. Lukierski, Z. Popowicr, P.
Presenajder, $q$-Deformations of Virasoro algebra and conformal
dimensions, {\it Phys. Lett. B}, {\bf 262}(1)(1991), 32--38.

\RE{CJ}R. Chakrabartit, R. Jagannathan, A $(p,q)$-deformed Virasoro
algebra, {\it J. Phys. A: Math. Gen.}, {\bf 25}(1992), 2607--2614.


\RE{CZ} T. L. Curtright, K. Zachos,  Deforming maps for quantum
algebras, {\it Phys. Lett.}, {\bf 243}(3) (1990), 237--244.


\RE{D1} V. G. Drinfel'd, Hopf algebras and the quantum Yang-Baxter
equation, {\it Soviet Math. Doklady}, {\bf 32}(1985), 254--258.


\RE{D2}V.G. Drinfel'd, Quantum groups, in: {\it Proceeding of the
International Congress of Mathematicians}, Vol.~1, 2, Berkeley,
Calif.~1986, Amer.~Math.~Soc., Providence, RI, 1987, pp.~798--820.


\RE{DG}D. W. Delius, M. D. Gould, Quantum Lie algebras, their
existence, uniqueness and $q$-antisymmetry, {\it Comm. Math. Phys.},
{\bf 185}(1997), 709--722.


\RE{FR} E. Frenkel, N. Reshetikhin, Quantum affine algebras and
deformations of the Virasoro and $W$-algebras, {\it Comm. Math.
Phys.}, {\bf 178}(1996), 237--264.

\RE{FLC}H. Fa, J. Li, Y. Cheng, Quantum group structure of the
$q$-deformed $W$ algebra $W_q$, arXiv:0803.0596v3.


\RE{GJP} S. Gao, C. Jiang, Y. Pei, The derivations, central
extensions and automorphism group of the Lie algebra $W$,
arXiv:0801.3911v1.

\RE{H} N. Hu, Quantum group structrue of the $q$-Deformed Virasoro
algebra, {\it Lett. Math. Phys.}, {\bf44} (1998), 99--103.

\RE{H2}N. Hu, $q$-Witt algebras, $q$-Virasoro algebra, $q$-Lie
algebras, $q$-holomorph structure and representations, {\it Colloq.
Alg.}, {\bf6}(1)(1999) 51--70.


\RE{HLS}J. T. Hartwig, D. Larsson, S. D. Silvestrov, Deformations of
Lie algebras using $\sigma$-derivation, {\it J. Alg.}, {\bf 295}
(2006), 314--361.



\RE{J1} M. Jimbo, A $q$-difference analogue of $U(g)$ and the
Yang-Baxter equation, {\it Lett. Math. Phys.},{\bf 10}(1985),
63--69.

\RE{J2} M. Jimbo, A $q$-difference analogue of $U(gl(N+1))$, Hecke
algebra and the Yang-Baxter equation,  {\it Lett. Math. Phys.}, {\bf
11}(1986), 247--252.

\RE{JP} W. Jiang, Y. Pei, On the structrue of Verma modules over the
$W$-algebra $W(2,2)$, {\it J. Math. Phys.}, {\bf 51}(2010), no.
022303, 8 pp.

\RE{K} C. Kassel, Cyclic homology of differential operators, the
Virasoro algebra and a $q$-analogue, {\it Comm. Math. Phys.}, {\bf
146}(1992), 343--356.


\RE{LSX}    J. Li, Y. Su, B. Xin, Lie bialgebras of a family of Lie algebras of Block type, {\it Chin. Ann. Math. Ser. B}, {\bf29} (2008), 487--500.


\RE{LZ} D. Liu, L. Zhu, Classification of Harish-Chandra modules
over the $W$-algebra $W(2,2)$, arXiv:0801.2601v2.

\RE{LGZ}D. Liu, S. Gao, L. Zhu, Classification of irreducible weight
modules over $W$-algebra $W(2,2)$, {\it J. Math. Phys.}
{\bf49}(2008), no. 113503, 6 pp.


\RE {L} Luu Thi Kim Thanh, Generalized $q$-deformation of Virasoro
algebra, {\it Comm. Phys.}, {\bf17}(4) (2007), 209--212.

\RE{MT} M. Mansour, E. H. Tahri, A $q$-deformation of Virasoro and
Kac-Moody algebras with Hopf structure, {\it Modern Phys. Letters},
{\bf14}(1999), 733--743.

\RE{OS}C.H. Oh, K. Singh, Realizations of the $q$-Heisenberg and
$q$-Virasoro algebras, {\it J. Phys. A: Math. Gen.}, {\bf 27}(1994),
3439--3444.


\RE{P} R.Parthasarathy., $q$-deformed W-algebras and the algebraic
structure of vertex operators in string theory, {\it Mod.Phys.Lett.
A}, {\bf 17}(2002), 399--411.


\RE{RM} M. A. Rego-Monteiroa, The quantum harmonic oscillator on a
circle and a deformed Heisenberg algebra, {\it Eur. Phys. J.} {\bf
C21}(2001), 749--756.

\RE{SW} A. Schmidt, H. Wachter, $q$-deformed quantum Lie algebras,
Arxiv-math. ph/0509032v1.

\RE{SS} G. Song, Y. Su, Lie bialgebras of generalized Witt type, {\it Science in China A}, 49(2006), 533--544.

\RE{SSX}    G. Song, Y. Su, B. Xin, Quantization of Hamiltonian-type Lie algebras,
{\it Pacific J. Math.}, {\bf240} (2009),  371--381.


\RE{SSW}    G. Song, Y. Su, Y. Wu, Quantization of generalized Virasoro-like algebras, {\it Linear Algebra Appl.}, {\bf428} (2008), 2888--2899.


\RE{WSS} Y. Wu, G. Song, Y. Su, Lie bialgebras of generalized Virasoro-like type, {\it Acta Mathematica Sinica-English Series,} {\bf22} (2006), 1915--1922.

\RE{WSS2} Y. Wu,  G. Song, Y. Su, Lie bialgebras of generalized Witt
type. II., {\it Comm. Algebra,} {\bf35} (2007), 1992--2007.

\RE{XSS} B. Xin, G. Song, Y. Su, Hamiltonian type Lie bialgebras, {\it Science in China A}, {\bf50} (2007), 1267--1279.

\RE{YaS}H. Yang, Y. Su, Lie super-bialgebra structures on the Ramond
N=2 super-virasoro algebra, {\it Chaos, Solitons and Fractals,}
{\bf40} (2009), 661--671.

\RE{YS} X. Yue, Y. Su, Lie bialgebra structures on Lie algebras of generalized Weyl type, {\it Comm. Algebra}, {\bf36} (2008), 1537--1549.

\RE{ZD} W. Zhang, C. Dong,  $W$-Algebra $W(2, 2)$ and the Vertex
Operator Algebra $L(\frac1 2, 0)\otimes L(\frac1 2, 0)$, {\it Comm.
Math. Phys.}, {\bf 285}(2009), 991--1004.
\end{thebibliography}
\end{document}